\RequirePackage{fix-cm}
\documentclass{article}

\usepackage{arxiv}
\usepackage[tight,footnotesize]{subfigure}
\usepackage{graphicx}
\usepackage{multicol}
\usepackage{amssymb,amsmath}
\usepackage[]{algorithm2e}
\usepackage{color}
\usepackage{soul}
\usepackage[normalem]{ulem}
\graphicspath{{figures/}}
\usepackage{pgffor}
\usepackage{hyperref}    

\begin{document}

\title{Machine Learning Interpretability and Its Impact on Smart Campus Projects}


\author{ Raghad~Zenki\\
	University of Northampton\\
	Northampton, United Kingdom \\
	\texttt{raghad.zenki@northampton.ac.uk} \\
	\And
	\href{https://orcid.org/0000-0003-1931-7959}{\includegraphics[scale=0.06]{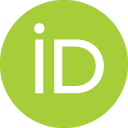}\hspace{1mm}Mu~Mu}\\
	University of Northampton\\
	Northampton, United Kingdom \\
	\texttt{mumu@ieee.org} \\
	}
	

\maketitle

\begin{abstract}
Machine learning (ML) has shown increasing abilities for predictive analytics over the last decades.  It is becoming ubiquitous in different fields, such as healthcare, criminal justice, finance and smart city. For instance, the University of Northampton is building a smart system with multiple layers of IoT and software-defined networks (SDN) on its new Waterside Campus. The system can be used to optimize smart buildings energy efficiency, improve the health and safety of its tenants and visitors, assist crowd management and way-finding, and improve the Internet connectivity.
 
With ML systems growing popular, many questions start to emerge: How much should human trust ML models? Will ML work on new “unknown” data? What can it tell about the world? Can human depend on AI partners? It is believed that the remedy of the concerns above hinges on the ability of ML/AI models to reason and interact with humans. Thus there is a need to develop models that can deliver good (performance) and explainable (interpretability) outcomes.

\end{abstract}

\section{The Definition and Impact of ML Interpretability}
\label{sec:intro}

Defining interpretability is a challenging task. Many researchers produce diverse motivations over the notion of interpretability and offer a myriad of features to classify a model as interpretable. However few have clearly and distinctly defined interpretability, especially its beneficiaries, shareholders, and overall significance to businesses and communities. Consequently, we can conclude that either: a) the interpretability is something globally agreed upon without being explicitly framed in words, or b) it is an inconspicuous term, and therefore, some explanations regarding the interpretability of ML models may lead to quasi-scientific claims. The author's observations of the literature imply that the latter seems to be the case. The variance and the occasional conflict, between the purpose of the interpretability and the explanation of many interpretable models, indicate that interpretability is not a monolithic concept, and it reflects a different number of diverse notions. 

Linguistically, Interpret means to explain a meaning or to present in understandable terms.  According to ML systems, interpretability could be defined as the capability to elaborate and reason the models' behaviour in a way that humans can comprehend. Nevertheless, the precise outlines of interpretation remain elusive; [1] described the explanation, in the context of psychology, as
``the currency in which human exchanged beliefs". The questions then arise: 
\begin{itemize}
\item What shapes an explanation? 
\item What makes an explanation understandable?
\item When/how it has to be generated? 
\item How ML applications interact with humans for informed decision marking?
\end{itemize}

Some research has characterised the interpretability as a means to give some sense of mechanism [2][3]. 

The potential influence of interpretability is multi-faceted. Interpretability is employed as a means to assert the raison d'etre of ML systems. Many requirements should be optimised in any ML systems, concepts of Fairness and Quality, reflect that the system is not discriminate against certain groups. Privacy would prove that confidential information in the data set is adequately protected. Other features such as Rigor and Robustness refer to the algorithm performance efficiency regarding the divergence of the input parameters. Causality is the ability to predict that the output will change due to some disturbance in the real system. Usability is about providing information that may help the users to complete a specific task. Lastly, Trust implies that a particular system has gained human confidence, such as aircraft collision avoidance systems. Some would argue that the research communities have classified some features such as privacy [4][5] and Reliability [6], and these formalisations followed by many rigorous studies in these areas with absolutely no need for further interpretations. Nevertheless, according to [7] ``Explanations may highlight the incompleteness", and the claim here is that interpretability assure that the ML desiderata, as mentioned above, are equally satisfied [8].

\section{Smart Campus Survey}

\begin{figure}[!htbp]
\centering
  \includegraphics[width=0.4\columnwidth]{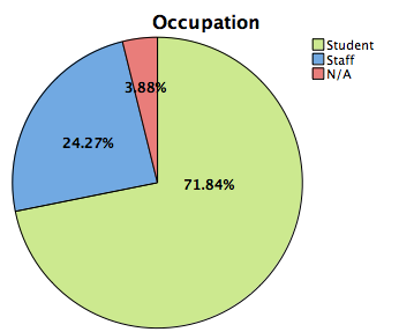}
  \caption{Participants’ occupations}\label{fig:1}
\end{figure}

Taking the notion of "Interpretability can instigate the users' trust" as a starting point, the author had conducted a survey (see Appendix) at the University of Northampton Waterside Campus, where many staff and students participated in it  (Fig. \ref{fig:1}). The purpose of the survey is to help us understand the public perception of data collecting and AI on a university campus in comparison with social media.  Furthermore, we investigate how ML interpretability could reshape the general opinion about data gathering and analysis.

\begin{figure}[!htbp]
\centering
  \includegraphics[width=0.9\columnwidth]{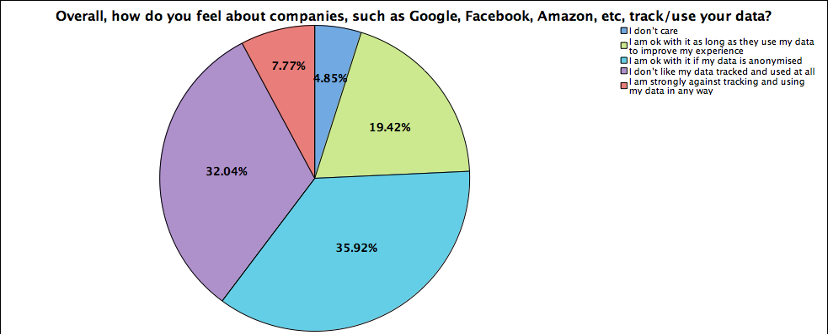}
  \caption{The participants’ response to their data being collected on Social Media}\label{fig:2}
\end{figure}

We collected the users' response to the fact that their data is being collected and analysed, on a daily basis, by companies such as Google, Facebook, and Amazon (Fig. \ref{fig:2}). 

\begin{figure}[!htbp]
\centering
  \includegraphics[width=0.9\columnwidth]{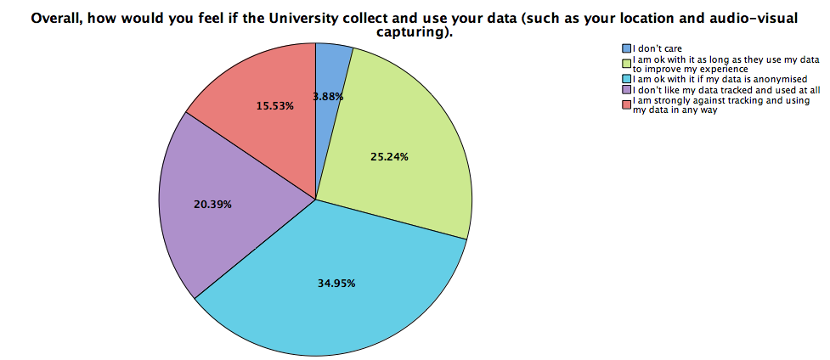}
  \caption{The participants’ response to their data being collected by the University}\label{fig:3}
\end{figure}

Next, we asked how they would feel about their data being captured and used by the university to improve their own experience on campus (Fig.\ref{fig:3}).

\begin{figure}[!htbp]
\centering
  \includegraphics[width=0.9\columnwidth]{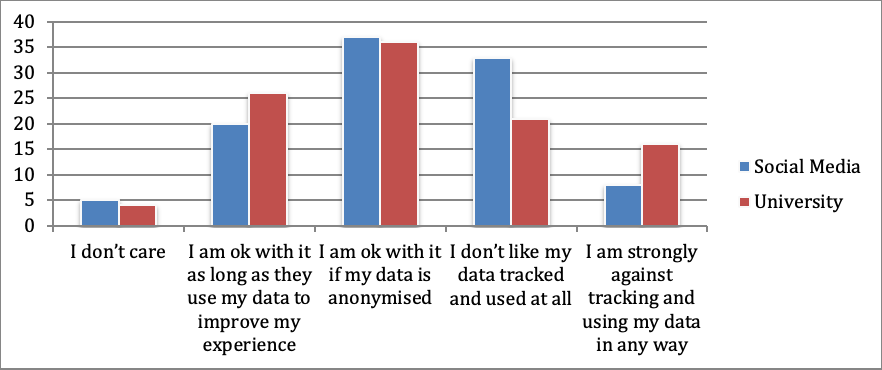}
  \caption{The participants’ response to their data being collected by the University vs. Social Media}\label{fig:4}
\end{figure}

Then we compared the results to verify our hypothesis, in which we assumed that the staff and students would trust the university more than social media. Our data shows that 64\% of the participants are okay with their data being used by the university, which is slightly higher than their response to social media (60\%). To our surprise, the results were close which means that students and staff would trust the university and social media with their data almost equally (fig.\ref{fig:4}). 

\begin{figure}[!htbp]
\centering
  \includegraphics[width=0.9\columnwidth]{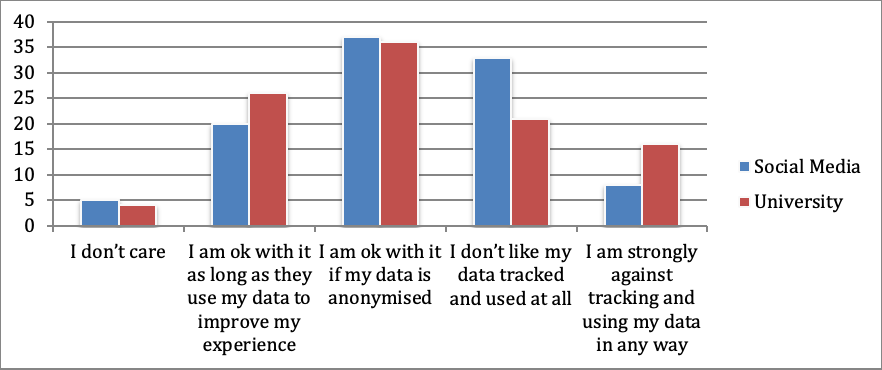}
  \caption{36\% of the participants who are against or strongly against their data being used response to using Interpretable ML model on campus}\label{fig:5}
\end{figure}

Then we focused on the people who are on the other side of the spectrum where there are 36\% of the participants against or strongly against their data being used in any format for any reason. When asked ``if the University’s Artificial Intelligence can explain to you how your data is used, how the decisions are made and let you make changes, would it change your opinion?”, over 45\% of this group of participants feel positive or very positive of the interpretable AI. (Fig.\ref{fig:5}). Looking at the opinions from all participants, around 55\% of the participants responded positive or very positive to ML interpretability.

\section{Conclusions and Future Work}

In conclusion, outlining the definition of the interpretability will lead to a better understanding of the fundamental purposes behind interpretable models. Consequentially, leading us to believe that interpretability in ML is the optimum approach to investigate, if we are to build an IoT-based model that can help the university understand how people use campus facilities, which leads to develop the university services, and enhance the students/staff experience on campus. Our initial data show that young university students find data gathering and AI-assisted smart campus generally acceptable while over half of the survey participants have positive or very positive opinions towards ML interpretability. Moreover, in future work, we are planning to build an interpretable smart system that reasons its decisions using information visualisation. Students will have the chance to provide feedback to the research team. In order to study how interpretability can bring positive changes to the perception of data-driven smart campus, we plan to conduct the survey again after deploying the system on campus.

\section*{Acknowledgements}
This work is supported by UK Research and Innovation (UKRI) under EPSRC Grant EP/P033202/1 (SDCN).

\section*{References}
\begin{enumerate}
\item Tania Lombrozo. The structure and function of explanations. Trends in cognitive sciences, 10(10): 464–470, 2006.
\item William Bechtel and Adele Abrahamsen. Explanation: A mechanist alternative. Studies in History and Philosophy of Science Part C: Studies in History and Philosophy of Biological and Biomedical Sciences, 2005
\item Nick Chater and Mike Oaksford. Speculations on human causal learning and reasoning. Information sampling and adaptive cognition, 2006.
\item incent Toubiana, Arvind Narayanan, Dan Boneh, Helen Nissenbaum, and Solon Barocas. Adnostic: Privacy preserving targeted advertising. 2010.
\item Cynthia Dwork, Moritz Hardt, Toniann Pitassi, Omer Reingold, and Richard Zemel. Fairness through awareness. In Innovations in Theoretical Computer Science Conference. ACM, 2012.
\item Moritz Hardt, Eric Price, and Nati Srebro. Equality of opportunity in supervised learning. In Advances in Neural Information Processing Systems, 2016.
\item Frank Keil, Leonid Rozenblit, and Candice Mills. What lies beneath? understanding the limits of understanding. Thinking and seeing: Visual metacognition in adults and children, 2004.
\item Doshi-Velez, F. and Kim, B. Towards A Rigorous Science of Interpretable Machine Learning. arXiv preprint arXiv:1702.08608v2, 2017. 
\end{enumerate}

\section*{Appendix}
\begin{figure}[!htbp]
\centering
  \includegraphics[width=0.73\columnwidth]{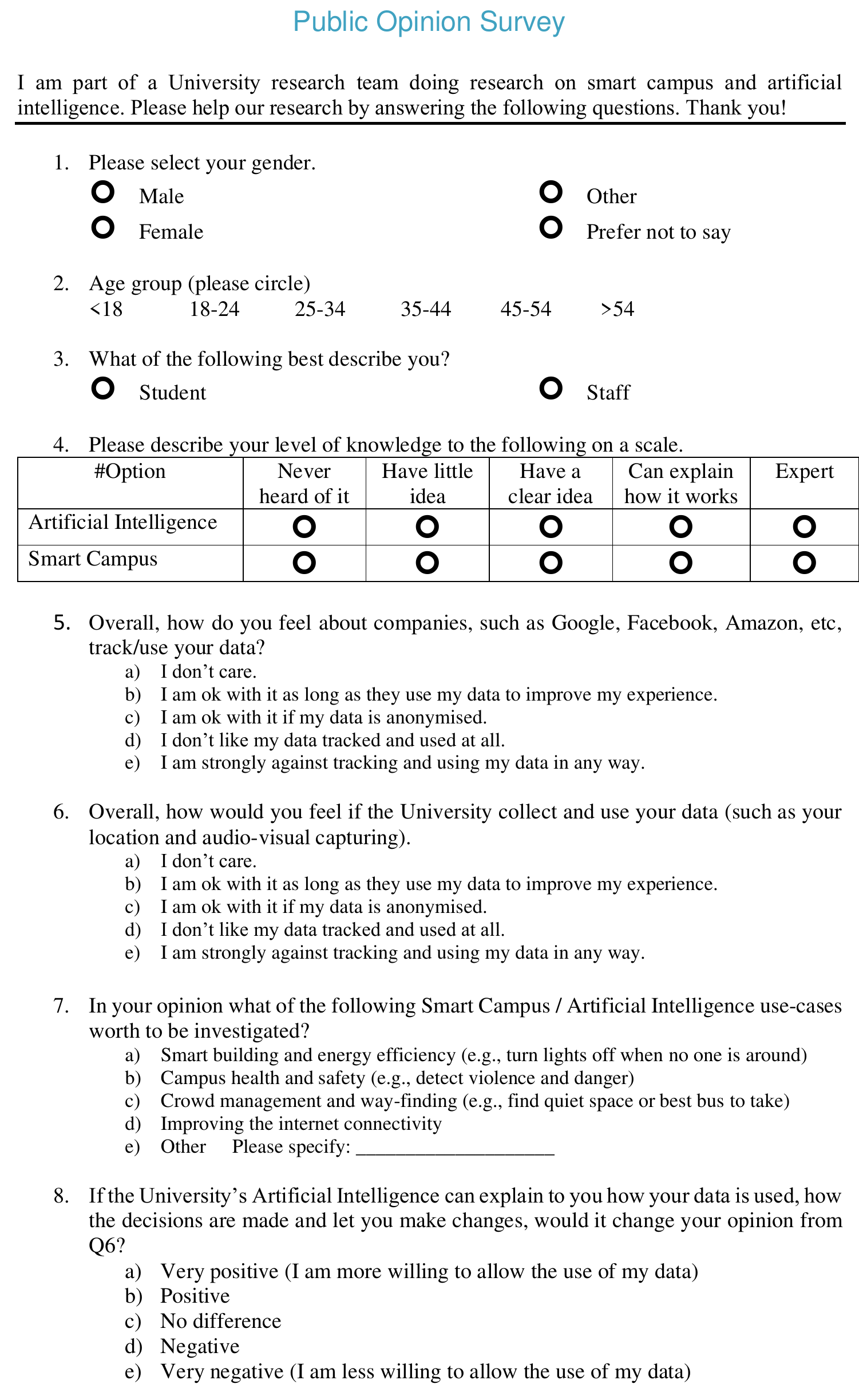}
  \caption{Questionnaire}\label{fig:6}
\end{figure}

\end{document}